\RequirePackage{lineno}
\documentclass[prb,twocolumn,amsmath,amssymb]{revtex4}

\usepackage{color}
\usepackage{graphicx}
\usepackage{dcolumn}
\usepackage{bm}

\def\U{URu$_2$Si$_2$~}

\def\be{\begin{equation}}
\def\ee{\end{equation}}
\def\bd{\begin{displaymath}}
\def\ed{\end{displaymath}}
\def\-{\phantom{-}}

\begin{document}

\title{Hidden order pseudogap in URu$_2$Si$_2$}

\author{J.T. Haraldsen$^{1,2}$}
\author{Y. Dubi$^3$}
\author{N.J. Curro$^4$}
\author{A.V. Balatsky$^{1,2}$}
\affiliation{$^1$Theoretical Division, Los Alamos National Laboratory, Los Alamos, NM 87545, USA}
\affiliation{$^2$Center for Integrated Nanotechnologies, Los Alamos National Laboratory, Los Alamos, NM 87545, USA}
\affiliation{$^3$School of Physics and Astronomy, Tel Aviv University, Tel Aviv 69978, Israel}
\affiliation{$^4$Department of Physics, University of California, Davis, California 95616, USA}

\begin{abstract}

Through an analysis and modeling of data from various experimental techniques, we present clear evidence for the presence of a hidden order pseudogap in \U in the temperature range between 25 K and 17.5 K. Considering fluctuations of the hidden order energy gap at the transition as the origin of the pseudogap, we evaluate the effects that gap fluctuations would produce on observables like tunneling conductance, neutron scattering and nuclear resonance, and relate them to the experimental findings. We show that the transition into hidden order phase is likely second order and is preceded by the onset  of non-coherent hidden order fluctuations.

\end{abstract}

\maketitle

\section{Introduction}

\U is a heavy fermion system that exhibits both magnetic and superconductive ordering. With a hidden order (HO) state at $T_{HO}$ = 17.5 K, \U provides a playground of physical phenomena that has intrigued condensed matter scientists for many years.\cite{mora:09,pine:97,hass:92,pals:85,schl:86,mapl:86,scho:87,hass:08} Over the past decades, experimental and theoretical investigations into \U have been centered around the mystery of the HO state.\cite{hass:92,baek:10,barz:95,oppe:10,shis:09,sikk:96,bonn:88,mapl:86,pals:85,schl:86,scho:87,hass:08} To explain this phase, various theories as to the origin of the complex magnetic and electronic states have provided different mechanisms ranging from magnetic helicity\cite{varm:06}, orbital magnetism\cite{chan:02}, octupolar ordering\cite{hanz:07}, unconventional spin density waves\cite{iked:98,mine:05}, and orbital hybridization\cite{dubi:11,pepin:11} just to name a few. While these theories attempt to explain the complexity of the HO phase, most only explain various elements and fail to reproduce all aspects of the state. This has made challenge posed by \U a stimulating topic and test-bed for new theories and experimental techniques.

Below $T_c = 1.5$ K,  \U enters an unconventional superconducting state.\cite{kasa:07} Recent point contact spectroscopy identified the presence of a pseudogap phase that precedes the superconducting transition up to 2 K and has also shown that the superconducting state does not exhibit standard superconducting characteristics.\cite{mora:09} While the non-BCS aspects of superconducting state are important, the existence of a HO phase has proved to be the most stimulating.
\begin{figure}[t]
\includegraphics[width=3.25in]{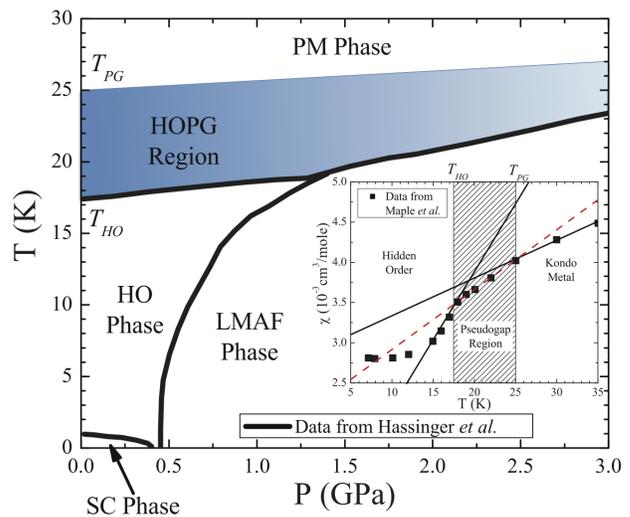}
\caption{(Color Online) The temperature versus pressure phase diagram for \U using data from Ref.~[\onlinecite{hass:08}] showing the paramagnetic (PM), hidden order (HO), superconducting (SC), and large moment antiferromagnetic (LMAF) phases. The shaded region denotes the proposed hidden order pseudogap (HOPG) region. The inset shows the susceptibility of \U as a function of temperature (observed by Maple $et~al.$\cite{mapl:86}), illustrating the existence of a pseudogap region. In the absence of a PG, the slopes of the susceptibility (solid black lines) would exhibit an abrupt change.  The discrepancies between the slopes (dashed red line) reveals the PG region (shaded area) between 17.5 and 25 K.}
\label{PG-PD}
\end{figure}

In this paper, we present evidence for a pseudogap (PG) crossover region before \U undergoes the phase transition into a HO state. However, in the previous analysis of the hidden order state, the existence of this precursory region has been largely overlooked. The pseudogap state may have been hinted at by Janik $et$ $al.$ when looking at itinerant spin excitations around the hidden order transition.\cite{jani:09} However, recent studies in far-infrared and point-contact spectroscopies have been discussing the region below 30K as a hybridization gap.\cite{leva:10,park:11} These claims however need to be discussed in the context of previous studies that have shown that a hybridization-like feature forms around 120 K.\cite{schm:10} Here, we present a more comprehensive discussion of the evidence for PG in the data, and provide theoretical simulations to analyze the effects of pseudogap within a mean-field model.  We re-analyze the experimental results and present theoretical calculations that supports the existence of a PG region below $T_{PG} \sim 25$ K. Using tunneling spectroscopy \cite{hass:92}, inelastic neutron scattering \cite{wieb:07}, and nuclear magnetic resonance (NMR) relaxation, we demonstrate how these quantities are affected by PG fluctuations. From the phase diagram of  URu$_2$Si$_2$, we infer that that PG region arises from fluctuations of the HO gap (see Fig.~\ref{PG-PD}) and could indicate competition with other gapped phase, like large moment antiferromagnetic phase at larger pressures. Large moment fluctuations are present at the  ambient pressure and therefore can provide the competing phase that produces PG features {\em in addition} to the HO phase that ultimately wins at lowest temperatures. In this regard, the scenario would be similar to a PG that can arise as a result of competition between magnetic order and superconductivity.

Our qualitative picture of a PG region consists of non-coherent order formed at a temperature $T_{PG} \sim 25 $ K, in our estimates, which is well above the transition temperature $T_{HO} = 17.5$ K.  One can make the general comment: a PG regime implies there is a precursor to the mean-field regime, where one has a fluctuating order parameter and gap developing above the transition yet with no true long-range order forming until one reaches $T_{HO}$. For the related discussion in recently discovered oxide and pnictide superconductors see Refs. \cite{taki:91,puch:96,renn:98,dama:03,ahil:08,klin:10}.

This phenomenon of fluctuating order and gap can occur due to amplitude or phase fluctuations. For the purposes of our analysis, we will focus on amplitude fluctuations. Yet it is possible that there are significant phase fluctuations of the HO order parameter.   Without the loss of generality, we will assume there are amplitude fluctuations of the gap associated with the HO state and these fluctuations would be a driving force for the PG behavior.

The central message of the paper is illustrated in the inset of Fig.~\ref{PG-PD}, where the PG can be easily observed in the magnetic susceptibility data from Ref.~\cite{mapl:86} and determined by a distinct change in slope preceding $T_{HO}$. While in the absence of a PG, one expects an abrupt change in the slope. Numerous probes that measure quasiparticle spectra can be used to reveal a pseudogap crossover. Using scanning tunneling microscopy (STM), the pseudogap is described as the opening of energy gap through a decrease in the density of states without the presence of coherence peaks,\cite{fisc:07} while nuclear magnetic resonance (NMR) has shown a distinct changes in the relaxation time within the $d$-wave cuprates.\cite{will:97} 

Our intent is to provide a qualitative phenomenological description of the fluctuations and is not meant to provide a comprehensive theoretical analysis of the HOPG state. In what follows, we discuss the evidence of the HOPG as observed in various experiments. We then illustrate how the effects of amplitude fluctuations would effect these measurements.

\begin{figure}
\includegraphics[width=3.5in]{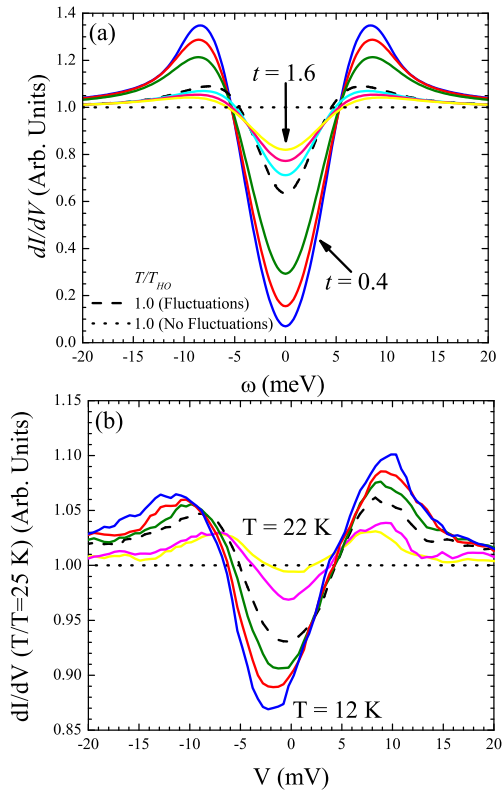}
\caption{(Color Online) (a) Simulated $dI/dV$ versus $\omega$ for $t$ = $T/T_{HO}$ = 0.4 to 1.6. Here, the pseudogap is shown for $t$ = 1.0 (dashed black) due to fluctuations within the energy gap. If no fluctuations occur, then the $t$ = 1.0 curve will be flat (dotted black). The suppression of the density of states at $T_{HO}$ demonstrates the presence of a pseudogap state. Here, $\sigma$ is given by $a$ = 3.0 and $b$ = 0.2.  (b) Normalized $dI/dV$ versus potential for \U from Ref. [\onlinecite{hass:92}]. We used normalization, where we divide the low temperature dI/dV(V,T) data by  high temperature data dI/dV(V,T = 25K).   We note that our  simple model does provide a reasonable comparison with the data, while we are not pursuing full fit of the data at this point. }
\label{G-HOPG}
\end{figure}

\section{Results and Discussion}

To investigate the pseudogap, we use three separate techniques to examining the qualitative effects of HO gap fluctuations. Within these simulations, we do not address the microscopic origins of the HO phase given the lack of a solid understanding of the HO. These calculation are meant to simply illustrate the effects of gap fluctuations for comparison to the experimental data. Since the pseudogap is likely to have a more microscopic origin, direct fitting of the order parameters is not necessary.

\subsection{Pseudogap in Point-Contact Spectroscopy}

Point contact spectroscopy (PCS) is a microscopic technique that provides a direct measurement of the density of states (DOS) for a material by examining its $I$-$V$ characteristics.\cite{fisc:07,duif:89,esud:94} By applying a voltage across a material the  tunneling conductance is directly related to the scattering of conduction electrons and the density of states.  If a system exhibits a pseudogap, one typically detects change within the DOS before the coherent ordering temperature.

PCS measurements on \U at various temperature that were performed almost two decades ago by Hasselbach $et~al.$ \cite{hass:92} demonstrated the presence of the HO state. However, Fig.~1 of Ref.~[\onlinecite{hass:92}] shows the measured conductance $G$  = $dI/dV$ as a function of bias voltage through a temperature range of 5.5 and 25 K. Analysis of the data shows the emergence of a coherent HO state at 17.5 K. From the data, it is clear that there is a distinct ``dip" in the density of states (DOS) between 22 K and the HO transition temperature $T_{HO}$. The lowering of the DOS before the transition is an indicator of pre-transition ordering without coherence in the system. Recent soft PCS measurements could be indicative of the same feature.\cite{lu:11}.

To model the effective HOPG, we assume the non-coherence can be modeled by fluctuations of the gap amplitude. Since the conductance $G$ is proportional to the DOS $\rho$, we can write the conductance as

\be
\displaystyle \frac{dI}{dV}  = \frac{\int \rho_0(\varepsilon,T)P(\Delta,T)\left(\frac{df}{d\varepsilon}\Big|_{\varepsilon\rightarrow\omega}\right)d\varepsilon d\Delta}{\int P(\Delta,T)d\Delta},
 \label{G}
\ee

where the DOS  is modeled as a general order parameter which induces a gap in the energy spectrum that is symmetric near Fermi energy for simplicity,

\be
\rho_0(\varepsilon,T) = \frac{|\omega|}{\sqrt{\omega^2-\Delta^2}} \label{DOS}
\ee

and we assume Gaussian distribution for amplitude fluctuations,

\be P(\Delta,T)=\frac{1}{\sigma\sqrt{2\pi}}e^{\frac{-(\Delta - \Delta_0(T))^2}{2\sigma^2}}~~.\label{Gaussian}
\ee

In Eq.~(\ref{G}), $f$ being the Fermi function, $\displaystyle f = (e^{-\varepsilon/k_BT}+1)^{-1}$
and $\sigma$ defines the extent of the order parameter fluctuations around its mean-field value $\Delta_0$. We take $\sigma$ to be a temperature dependent function $\sigma$ = $ae^{-b\sqrt{(1-t)^2}}$, where $a$ produces fluctuations and $b$ introduces a temperature dependence and both are positive constants. The temperature dependence of the mean-field gap $\Delta_0(T)$ is defined in the typical way,

\be
\Delta_0(T) =
\begin{cases}
\frac{\Delta_{gap}\sqrt{1-t}}{1-\frac{t}{2}}, & 0 \le t < 1 \\
0, &  t \ge 1
\end{cases}\\
\label{gap}
\ee

where $t$ = $T/T_{HO}$ and we set $\Delta_{gap} \approx$ 5.0 meV to comply with available data on \U.

Figure \ref{G-HOPG} shows the conductance as a function of $\omega$ for $t$ ranging from 0.4 to 1.6, calculated using Eq.~(\ref{G}). The dashed black line denotes $t$ = 1.0, where $T$ = $T_{HO}$. For the case of no PG, the $t$ = 1.0 line should be flat (dotted black line) indicating no fluctuations. However, due to the gap fluctuations, the simulation demonstrates a suppression of the DOS even at the critical temperature $T_{HO}$, which mimics the behavior of the PG observed in experiment.
\begin{figure}
\includegraphics[width=3.75in]{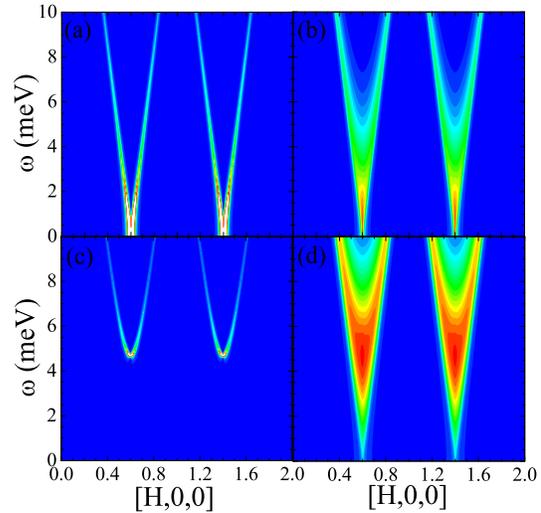}
\caption{(Color Online) Spin susceptibility as a function of energy and momentum. (a) and (b) show the gapless system ($t\ge1.0$) without and with gap fluctuations, respectively. (c) and (d) show the gapped system ($t$ = 0.5) without and with fluctuations, respectively. The clear broadening of the spin excitations denotes a precursory PG to the HO phase.}
\label{SS-HOPG}
\end{figure}

\subsection{Pseudogap in Inelastic Neutron Scattering}

Through the use of inelastic neutron scattering (INS), Wiebe $et~al.$ investigated the spin excitations above and below $T_{HO}$\cite{wieb:07} and revealed itinerant-like spin excitations at incommensurate wavevectors ($H$ $\sim$ (0.6,0,0) and (1.4,0,0)) at about 5 meV ($\Delta_0$) as well as lower energy commensurate spin excitations. Both features have been the motivation for the recent hybridization wave and other proposals for HO models (see, e.g. \cite{dubi:11,pepin:11,varm:06,chan:02,hanz:07,iked:98,mine:05}).

By evaluating the data, we conclude that inelastic neutron scattering data also signal the existence of the HOPG at 20 K. This is deduced from the appearance of distinct broadening of the spin excitations at 20 K, which can be explained by gap fluctuations of a PG as HO phase begins to order (see Fig.~2 of Ref. [\onlinecite{wieb:07}]).

To illustrate the effects of a PG on these features, we add similar gap fluctuations to the calculation of the spin susceptibility as a function of $\omega$ and $q$. The spin susceptibility is therefore defined as

\be
\displaystyle \chi^{\prime\prime}(\omega,q)  = \int_0^{\infty}{\rm Im}\left[\frac{P(\Delta,T)}{\omega^2-\omega_q^2+i\delta}\right]d\Delta
\label{SS}
\ee

where $\omega_q$ = $\sqrt{(cq)^2+\Delta^2}$, $c$ describes the spin-wave velocities ($\sim$45 meV \AA),  and $\delta$ is a small broadening (note that without the integration over $\Delta$ one obtains the usual expression for the spin susceptibility).

\begin{figure}
\includegraphics[width=3.5in]{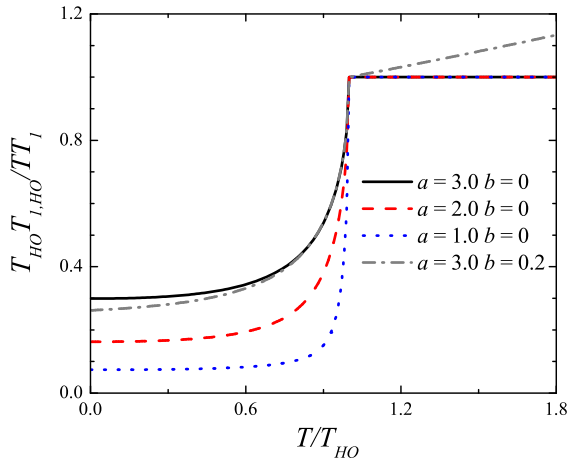}
\caption{(Color Online) The temperature dependence of the NMR relaxation time for gap fluctuations with and without a temperature dependent distributions calculated from integrating Eq.~\ref{NMR} over the fluctuations of $\Delta$. The increased presence of gap fluctuations increases the relaxation time and softens the gap suppression. The addition of a temperature dependent $\sigma$ ($b$ = 0.2) produces a sloped response for $t >$ 1.0 (dash-dotted gray).}
\label{NMR-fig}
\end{figure}

Figure \ref{SS-HOPG}(a)-(d) shows the spin susceptibility (calculated from Eq.~(\ref{SS}) with the HO gap and fluctuations defined by Eq.~(\ref{DOS}-\ref{Gaussian}), as function of $\omega$ and $q$ at $t$ = 1.0 ((a) and (b)) and $t$ = 0.5 ((c) and (d))(see Eq. \ref{gap}). The left panels (a) and (c) do not include any gap fluctuation (i.e. no integration over $\Delta$), while the right panels (b) and (d) are simulated with gap fluctuation similar to the point contact experiment.

The introduction of gap fluctuations produce a widening of the excitations, which allows suppression of spectral weight to be observed above the transition temperature. Experimentally, the generalized broadening shown in Figs. (1-3) of Ref.~[\onlinecite{wieb:07}] demonstrates the existence of gap fluctuations and ordering above the transition temperature and denotes the presence of a PG before the HO phase transition.

\subsection{Pseudogap in Nuclear Magnetic Resonance}

Through the use of NMR, one can probe the spin-lattice relaxation rate and determine the existence of an ordered phase transition.\cite{slic:80} By introducing fluctuations to the HO energy gap, it is expected that multiple or varying lines shapes will indicate a PG regime. To model this probe, we examine the effects of the gap fluctuations on the NMR relaxation rate $(TT_1)^{-1}$, which is given by

\be
\frac{1}{TT_1} = \frac{\gamma_n^2k_B}{2\mu_B^2}\sum_q \frac{q^2}{2\pi^2} \frac{\chi^{\prime\prime}(\omega,q)}{\omega} \Big|_{\omega \rightarrow 0},
\label{NMR}
\ee

where $\gamma_n$ is the nuclear gyromagnetic ratio, $k_B$ is Boltzmann's constant, and $\mu_B$ is the Bohr magneton.\cite{slic:80}

In Fig.~\ref{NMR-fig}, the NMR relaxation time as a function of temperature is calculated from integrating Eq.~(\ref{NMR}) over the fluctuations of $\Delta$ (Eq.~(\ref{Gaussian})), for different values of order parameter fluctuation range $a$=1 (blue dotted), 2 (red dashed), 3 (black solid) meV with $b$ = 0. The dash-dotted (gray) curve is for $a$ =  3.0 meV and $b$ = 0.2 to demonstrate the effect of a temperature dependent $\sigma$.

As shown in Fig.~\ref{NMR-fig}, when gap fluctuations are small ($a \le$ 1.0), the relaxation time shows a dramatic decrease at the transition temperature. The drop is caused by the opening of an energy gap at the Fermi level. However, as gap fluctuations are increased ($a >$ 1.0), the presence of non-coherent order helps broaden in the relaxation time suppression due to an increase in the DOS. A similar broadening effect has been observed demonstrated in multiple NMR investigations of URu$_2$Si$_2$,\cite{mats:96,koho:96} which indicates the clear presence of order before the transition.

Recent NMR measurements\cite{curr:11} have observed a slight rounding of the transition boundary and a sloped response for $t >$ 1.0. The rounding of the transition edge has also been observed, for instance, in NMR studies of YBCO. \cite{taki:91} This seems to indicate that $\sigma$ is temperature dependent. In Fig.~\ref{NMR-fig}, the dash-dotted (gray) line shows the effect of a small temperature dependence on the fluctuation distribution ($b$ = 0.2). This temperature dependence has minimal effect on the other probes, but produces the linear response to the NMR relaxation time. Future measurements and investigations will be able to clarify this point.

\section{Conclusion}

We demonstrate that there is clear evidence for a HOPG in various observations of URu$_2$Si$_2$ around $T_{PG} \sim 25$  K. We simulate the effect of a PG in multiple experimental probes through the introduction of fluctuations of the HO gap. Through a comparison of our simulations and various experiment observations, we conclude there is a significant evidence for presence of a HOPG. In addition to the probes discussed here, we point to other techniques (heat capacity, resistivity, optical conductivity etc.) which also hint to the presence of the HOPG state, although this evidence is not as clear.

The richness of debate around the nature of HO and normal state from which it emerges, emphasizes the importance of investigating this region fully. Results analyzed here and discussions on the HOPG pose additional constraints on the models for both normal state and HO and will lead to better understanding of the origin and nature of the HO phase.

\section{Acknowlegdements}

We would like to acknowledge useful discussions with M. Graf, J. J. Su, D. Parker, T. Durakewicz, T. Timusk, T. Das. This work was supported, in part, by UCOP-TR01, by  the Center for Integrated Nanotechnologies, a U.S. Department of Energy, Office of Basic Energy Sciences user facility and in part by the LDRD.  Los Alamos National Laboratory, an affirmative action equal opportunity employer, is operated by Los Alamos National Security, LLC, for the National Nuclear Security Administration of the U.S. Department of Energy under contract DE-AC52-06NA25396.

\end{document}